\providecommand{\exclude}[1]{}

\documentclass[10pt, aps, pra, superscriptaddress, nofootinbib,
               amsmath, 
               twocolumn, preprintnumbers,
               showpacs, 
               raggedbottom,
               floatfix,
               longbibliography, 
               colorlinks=true, linkcolor=blue, citecolor=blue, urlcolor=blue,
              ]{revtex4-1}
\pdfoutput=1
\usepackage[T1]{fontenc}
\usepackage[utf8]{inputenc}
\usepackage[caption=false]{subfig}

\usepackage[secnoindent,
            ]{my_paper} 
\usepackage{my_acronyms}        
\usepackage[english]{babel}     
\usepackage{adjustbox}          
\usepackage{colortbl}
\usepackage{xparse}             

\usepackage[
            tightpage]{preview}

\usepackage{cleveref}
\usepackage{xcolor}

\renewcommand{\vect}[1]{\boldsymbol{#1}}

\begin{document}
\newlength{\mybaselineskip}
\setlength{\mybaselineskip}{\baselineskip}

\title{Detecting Entrainment in Fermi-Bose Mixtures}

\author{Khalid Hossain}%
\email{mdkhalid.hossain@wsu.edu}%
\affiliation{Department of Physics \& Astronomy, %
  Washington State University, Pullman, Washington 99164--2814, USA}%

\author{Subhadeep Gupta}%
\email{deepg@uw.edu}%
\affiliation{Department of Physics, %
  University of Washington, Seattle, Washington 98195--1560, USA}

\author{Michael McNeil Forbes}%
\email{mforbes@alum.mit.edu}%
\affiliation{Department of Physics \& Astronomy, %
  Washington State University, Pullman, Washington 99164--2814, USA}%
\affiliation{Department of Physics, %
  University of Washington, Seattle, Washington 98195--1560, USA}

\date{\today}
\preprint{}


\begin{abstract}

We propose an experimental protocol to directly detect the Andreev-Bashkin effect (entrainment) in the bulk mixture of a bosonic and fermionic superfluid using a ring geometry.
Our protocol involves the interferometric detection of the entrainment-induced phase gradient across a superfluid due to the flow of another in which it is immersed. The choice of ring geometry eliminates variations in the stronger mean-field interaction which can thwart the detection of entrainment in other geometries.
A significant enhancement of the entrainment phase shift signal is possible, if the dimer-boson scattering length turns out to be large, which can be measured by tuning the interaction to the limit of miscibility of the two superfluids.
With suggested improvements and careful design implementation, one may achieve $\approx 67$\% shift in the interferometer fringes.

\end{abstract}

\maketitle


\section{Introduction}
Superfluid entrainment plays a crucial role in the dynamics of superfluid mixtures in neutron stars.
The effect -- whereby a flowing superfluid drags along or ``entrains'' another superfluid despite the lack of dissipation -- was first predicted by Andreev and Bashkin for mixtures of superfluid \ce{^{3}He} and \ce{^{4}He}~\cite{Andreev-bashkin-three-velocity, Volovik:1975a}.
Since its prediction, superfluid entrainment has been studied extensively in the contexts of neutron stars, nuclear physics~\cite{chamel17, Haskell:2018, antonelli17, leinson17, Bulgac:2018}, and cold atoms~\cite{nespolo17, Parisi:2018, chevy15, kravchenko08, fil04:2, fil04, fil05, Demin2002}, but despite almost half a century of study, it has yet to be directly observed in experiments.

Cold atom systems provide a new environment in which to search for entrainment where one has exquisite control over both geometry and interactions~\cite{Cheng-feshbach-review}.
Entrainment manifests as an effective potential for one superfluid species induced by the flow of another. This property can be used to induce a phase winding which can then be measured using interferometry. In this paper, we outline a procedure using this effect to directly observe entrainment in a two species mixture of superfluid \ce{^{6}Li} and \ce{^{174}Yb}, taking the advantage of the large mass of the bosons and the availability of species specific potentials.

On a nuclear scale, neutron stars are cold and the crust is expected to comprise a mixture of superfluid neutrons and superconducting protons~\cite{chamel:viewpoint2011}.
These superfluids act as an internal reservoir of angular momentum, which may be suddenly transferred to the crust through vortex (un)pinning or hydrodynamic instabilities, resulting in a sudden increase in the rotation rate of the stars, observed as a ``glitch'' in pulsar data~\cite{Anderson:1975, Alpar:1977, Jones:1991, Link:1991, Link:1999, Mannarelli:2007bs, Pizzochero:2011, Andersson:2011, Andersson:2012, Warszawski:2011, Grill:2012, Warszawski:2012, Melatos:2014, Eysden:2013, Lasky:2013, Chamel:2013a, Seveso:2016, Archibald:2013, Lyne:2014, Haskell:2015}.
Although the detailed mechanism behind glitches remains a mystery, it is clear that superfluid dynamics and entrainment between the proton and neutron fluids play a significant role~\cite{Babaev} owing to the strength of the nuclear interactions.
Unfortunately, being light-years away, astrophysical bodies like neutron stars and pulsars are extremely difficult to measure.

For this reason we turn to ultracold atom experiments, where one can create superfluid mixtures~\cite{Modugno-Two-Superfluid, Ferrier-Barbut-Bose-Fermi, Deep-two-element} to act as quantum simulators of neutron-star physics~\cite{Truscott2570, graber17}.
The challenge with terrestrial experiments is that the magnitude of the entrainment effect depends on the inter-species interaction strength $g_{ab}$, which must generally be small $g_{ab}^2 < g_{aa}g_{bb}$ to allow the superfluids to mix.
This frustrates detection in \ce{^{3}He} and \ce{^{4}He} superfluid mixtures~\cite{Pickett2002} where the mutual interactions require temperatures below that which can be achieved by state-of-the-art cryogenics in order for the fluids to not phase separate~\cite{Salmela-Effective-Interaction, Riekki-Adiabatic-Melting}.
In dilute Bose gases, the entrainment is further suppressed as it occurs at second-order, depending on the square of the inter-species interactions~\cite{fil04:2, shevchenko07}.
This thwarts attempts to measure the entrainment through, for example, \@ modifications of the dipole frequency~\cite{Stringari-Counter-Flow} because mean-field effects dominate.
For the \ce{^{174}Yb}-\ce{^{6}Li}- superfluid mixture considered here, the calculated value of the dipole frequency shift due to entrainment is only $\approx 0.02 \%$, two orders of magnitude smaller than the observed shift~\cite{Deep-two-element} from mean-field effects.

In addition to serving as a suitable system to observe entrainment, a mixture of fermionic and bosonic superfluids also serves as a versatile platform for other studies in many-body physics. The \gls{FFLO} phase~\cite{Chandrashekhar-clogston, Liao:2010, Partridge503:2006}, time-reversal-invariant superfluids with exotic topological properties~\cite{Time-reversal-pra}, rotational responses~\cite{gren2020rotational} and the dynamics and structure of solitons~\cite{Dark-Bright-njp}, can be explored by tuning the mass-ratios, particle densities, and interparticle interactions. Interesting physical properties such as, the ground state characteristics of a mixture~\cite{Ol-bosonic-fermionic-atoms}, quasiparticle excitation spectrum~\cite{Ol-quasiparticle-spectrum}, general phase diagram~\cite{Ol-phase-diagram}, phase competition among density wave orderings and superfluid pairings~\cite{Ol-competing-orders}, and many-body effects in the mixture~\cite{Ol-many-body-effects} have been explored in great detail in the presence of optical lattice potentials, facilitating the interplay between non-linearity and periodicity. Entrainment, in particular, has been investigated theoretically for two-\cite{Ol-drag-two-babaev, Ol-drag-two-hofer, Ol-drag-two-linder, Contessi:2021} and multi-component~\cite{Ol-drag-multi-hartman} Bose-Bose mixture in optical lattices.
By introducing an optical lattice to the superfluid mixture, one may enhance the entrainment signal by softening the phonon dispersion relationship~\cite{Ol-drag-two-hofer}.

A mixture of superfluid \ce{^{6}Li} and \ce{^{174}Yb} provides a favorable environment in which to detect entrainment.
The \ce{^{6}Li} scattering length may be tuned with a Feshbach resonance~\cite{schunck05}, allowing us to amplify entrainment effects while ensuring that the fermionic \ce{^{6}Li} superfluid remain mixed with the bosonic superfluid \ce{^{174}Yb}.
As mentioned above, although dipole oscillations provide a natural place to search for entrainment, mean-field effects dominate the frequency shift in cold atom systems.
We shall use these mean-field effects to test our microscopic theory, but then carefully design a procedure to eliminate these mean-field effects by placing the two components in ring traps.
Our proposal is similar to that suggested by~\cite{shevchenko07}, but differs in details for removing mean-field contamination and in using interferometry to measure the effective entrainment. Another idea proposed in~\cite{nespolo17} is to enhance entrainment with tight confinement to induce a dimensional reduction. In contrast, with the enhancements discussed in this paper, we expect to be able to directly observe bulk entrainment in 3D.

The rest of the paper is organized as follows. In section~\ref{sec 2}, we derive the mean-field equations of motion for the mixture with entrainment interaction.
Section~\ref{sec 3} discusses the proposed experiment.
In section~\ref{sec 4} we estimate the phase shift.
The main results and detail discussion of the physics of the detection is presented in section~\ref{sec 5}: suitable experimental parameters (\ref{sec 5a}) and characteristics of the detectable signal (\ref{sec 5b}).
Finally, we conclude in section~\ref{sec 6}.

\section{Theory}\label{sec 2}

First we consider a dilute single-component bosonic superfluid at $T=0$.
If the interactions are weak, we may describe this in the mean-field approximation by the usual \gls{GPE} for the condensate wavefunction $\psi_{b}$.
This is related through a Madelung transform to the superfluid density $n(\vect{r}, t)$ and phase velocity $\vect{v} = \frac{\hbar}{m}\vect{\nabla}\phi$ where $\phi$ is the momentum potential and $\psi_{b} = \sqrt{n_{b}}e^{\I \phi}$.
The \gls{GPE} follows from the following Lagrangian density:
\begin{subequations}
  \begin{gather}\label{eq:lagrangian1}
    \mathcal{L}[\psi_{b}] = \I \hbar \psi_{b}^{\dagger}\dot\psi_{b} - \mathcal{E}(\psi_{b}),\\
    \mathcal{E}(\psi_{b}) = \frac{\hbar^2\vect\nabla\psi_{b}^{\dagger}\cdot\vect\nabla\psi_{b}}{2m_{b}} + \overbrace{\frac{g_{bb}}{2} \left(\psi_{b}^{\dagger}\psi_{b}\right)^2}^{\mathcal{E}_{b}(n_{b})},
  \end{gather}
\end{subequations}
where $g_{bb} = 4\pi \hbar^2 a_{bb}/m_{b}$ is the coupling constant, $a_{bb}$ is the $s$-wave scattering length between the bosons, and $m_{b}$ is the mass of the bosons.
The equations of motion give rise to the usual \gls{GPE}:
\begin{equation}
  \I \hbar \dot\psi_{b}(\vect{r}, t) = -\frac{\hbar^2 \nabla^2}{2 m_{b}}\psi_{b}(\vect{r}, t) + \overbrace{g_{bb}n_{b}(\vect{r}, t)}^{\mathcal{E}_{b}'(n_b)}\psi_{b}(\vect{r}, t).
\end{equation}
This formulation describes the dynamics of a weakly interacting \gls{BEC} where the the equation of state for homogeneous matter is $\mathcal{E}_{b}(n_{b}) = g_{bb}n_{b}^2/2$.

However, the same formulation can be modified to describe the \gls{BEC} limit of the \gls{BCS}-\gls{BEC} crossover of symmetric (unpolarized) fermionic superfluids with resonant $s$-wave interactions (see~\cite{Zwerger:2011} for a review).
In this limit, the fermionic superfluid can be modeled with a similar equation, known as an \gls{ETF} model, which describes the fermionic superfluid as a gas of condensed bosonic-dimers with number density $n_D = n_{f}/2$.
The \gls{ETF} for this system is similar to the \gls{GPE} with three modifications:
1) the mass is replaced by the dimer mass $m_D = 2m_{f}$ where $m_{f}$ is the mass of the fermionic components (\ce{^6Li} in our case),
2) the density $n_D = \psi_D^{\dagger} \psi_D = n_{f}/2$ is interpreted as the dimer density rather than the fermion density, and
3) the interaction is replaced by the homogeneous equation-of-state in the crossover $\mathcal{E}_{f}(n_{f})$ that depends on the magnetic field $B$ through the fermion-fermion scattering length $a_{ff}(B)$:
\begin{gather}
  \I \hbar \dot\psi_D(\vect{r}, t) = -\frac{\hbar^2 \nabla^2}{2 m_D}\psi_D(\vect{r}, t) + \pdiff{\mathcal{E}_{f}(2n_D)}{n_D}\psi_D(\vect{r}, t).
\end{gather}

Well justified in the \gls{BEC} limit, this formulation continues to work quite well in the unitary limit where $a_{ff} \rightarrow \infty$, where it correctly describes~\cite{Ancilotto:2012} the hydrodynamics of rather violent collisions~\cite{Joseph:2011} and qualitative properties of vortex dynamics~\cite{Forbes:2012b}.
Some of this success comes from the fact that this \gls{ETF} model correctly reproduces the low-lying phonon spectrum:
\begin{gather}\label{eq:EphFermi}
  E_{\mathrm{ph}}(\hbar k) = \sqrt{\frac{\hbar^2k^2}{4m_{f}}\left(\frac{\hbar^2k^2}{4m_{f}} + 4\mathcal{E}''_{f}(n_f)n_f\right)}, \qquad
  k \ll k_F,
\end{gather}
where $4\mathcal{E}''_{f}(n_f) = \partial^2\mathcal{E}_{f}(2n_{D})/\partial n_D^2$, although it fails to predict pair-breaking and associated phenomena which occur near $k \sim k_F$.
This correctly reproduces the speed of sound $c_s$ where $E_{\mathrm{ph}}(\hbar k) \sim \hbar k c_s$ is linear in both the \gls{BEC} and the \gls{UFG} limits:
\begin{subequations}
  \begin{align}
    c_s^2 &= \frac{g_{DD} n_D}{m_D},
    && \left.\mathcal{E}_{f} \sim \frac{-\hbar^2n_{f}}{2m_{f} a_{ff}^2} + \frac{g_{DD}n_D^2}{2}\right|_{a_{ff}\ll 1}, \tag{BEC}\\
    c_s^2 &= \frac{\xi}{3}v_{f}^2,
    && \left.\mathcal{E}_{f} \sim \xi \frac{\hbar^2 (3\pi^2 n_{f})^{5/3}}{10 \pi^2 m_{f}}\right|_{a_{ff}\rightarrow \infty}, \tag{UFG}
  \end{align}
\end{subequations}
where $g_{DD} \approx 0.6 a_{ff}$ is the dimer-dimer scattering length~\cite{Petrov-Fermion-Dimer}, and $\xi = \num{0.3742(5)}$~\cite{Forbes:2012} is the universal Bertsch parameter~\cite{Bertsch:1999:mbx, Baker:1999, baker00:_mbx_chall_compet}, combining experiment~\cite{Ku:2011, Zurn:2013} and \gls{QMC}~\cite{FGG:2010, Carlson:2011} values.
In our analysis, we use a Pad\'e approximant for $\mathcal{E}_{f}(n_{f})$ that correctly interpolates between the unitary and \gls{BEC} limits including \gls{QMC} values for the Tan contact~\cite{Drut:2011, Rossi:2018} and few-body parameters in the deep \gls{BEC} limit~\cite{Astrakharchik:2004, Adhikari:2009}.

The Lagrangian~\eqref{eq:lagrangian1} is Galilean invariant under a boost of velocity $\vect{v}$ with an appropriate phase redefinition:
\begin{equation}
  \psi(\vect{x}, t) \rightarrow e^{\I\phi} \psi(\vect{x} - \vect{v} t, t)
  \qquad \hbar \phi = m\vect{v} \cdot \vect{x} - \frac{mv^2t}{2}.
\end{equation}
The generalization for two components -- a boson $\psi_b$ and a dimer $\psi_D$ -- has a similar form, but admits other terms allowed by Galilean covariance.
For the weakly interacting case, we add a term coupling the phase gradients of the two components:
\begin{subequations}\label{eq:functional}
  \begin{gather}
    \psi_{D,b} = \sqrt{n_{D,b}}e^{\I \phi_{D,b}} \qquad m_{D,b}\vect{v}_{D,b} = \hbar \vect{\nabla} \phi_{D,b}\\
    \mathcal{L} = \I \hbar \big(\psi_D^{\dagger}\dot\psi_D + \psi_b^{\dagger}\dot\psi_b\big) - \mathcal{E}[\psi_D, \psi_b, \vect\nabla\psi_D, \vect\nabla\psi_b],\\
    \begin{multlined}\nonumber
      \mathcal{E}[\psi_D, \psi_b] = \frac{\hbar^2}{2m_D} \abs{\vect\nabla\psi_D}^2
      + \frac{\hbar^2}{2m_b} \abs{\vect{\nabla}\psi_b}^2 +\\
      - \underbrace{\rho_{\mathrm{dr}}(n_D, n_b)\abs{\vect {v}_D - \vect {v}_b}^2}_{\mathcal{E}_{\mathrm{ent}}}
      + \underbrace{g_{bb}\frac{n_b^2}{2} + g_{Db}n_Dn_b + \mathcal{E}_f(2n_D)}_{\mathcal{E}_h(n_D, n_b)},
    \end{multlined}\\
    g_{bb} = \frac{4 \pi \hbar^2 a_{bb}}{m_b}, \qquad
    g_{Db} = \frac{2 \pi \hbar^2 a_{Db}(m_D + m_b)}{m_Dm_b}.
  \end{gather}
  Here $\rho_{\mathrm{dr}}$ has the dimension of mass-density, as in $\rho = mn$.
$\mathcal{E}_{\mathrm{ent}}$ is the entrainment term:
  \begin{equation*}
    \abs{\vect{v}_D - \vect{v}_b}^2 = \frac{\hbar^2}{m_Dn_Dm_bn_b}\Biggl|\sqrt{\frac{m_b}{m_D}}\psi_b \vect{\nabla} \psi_D - \sqrt{\frac{m_D}{m_b}}\psi_D \vect{\nabla} \psi_b \Biggr|^2.
  \end{equation*}
\end{subequations}
This term is manifestly Galilean invariant since it contains a difference between what is sometimes called the ``superfluid velocities'' $\vect{v}_i = \hbar \vect{\nabla}\phi_i/m_i$.
Note: for simple superfluids with only a single quadratic gradient term $\abs{\vect{\nabla}\psi}^2$, this superfluid velocity corresponds with the group velocity $\vect{j}_i/m_in_i$, but the presence of additional gradient terms, such as this entrainment term, changes the relationship between the phase gradients $\vect{\nabla}\phi$ and the group velocitiy.
In particular, as we shall see, entrainment allows a phase gradient to appear in a system even in the absence of a current, providing a way to detection of entrainment through interference.

Miscibility of the bosonic and fermionic superfluids -- that they occupy the same physical space -- is a requirement for detecting entrainment.
A necessary condition for miscibility is that the energy density be convex:
\begin{multline}
  \left(\frac{\partial^2\mathcal{E}_h}{\partial n_D \partial n_b}\right)^2
  \frac{\partial^2\mathcal{E}_h}{\partial n_D^2}
  \frac{\partial^2\mathcal{E}_h}{\partial n_b^2}
   \implies\\
  \frac{\pi\hbar^2(1+\frac{m_{b}}{m_D})^2}{4m_{b}a_{bb}}
  <
  \frac{\mathcal{E}_f''(n_f)}{a_{Db}^2}.
  \label{eq:miscibility}
\end{multline}
For our setup, the quantities on the left hand side are fixed, but the right-hand-side can be adjusted using the \ce{Li} resonance through the $B$-field dependence of $\mathcal{E}_f(n_f)$ to ensure that our mixture remains miscible as we attempt to maximize the entrainment.

In particular, the miscibility depends sensitively on the dimer-boson scattering length $a_{Db}$.
Na{\"\i}vely one might expect this to be twice the Fermi-Bose scattering-length $a_{Db} \sim 2a_{fb}$, but mean-field effects enhance this $a_{Db} \sim 3.87a_{fb}$ for our $m_{\ce{Yb}}/m_{\ce{Li}} \approx 29$ mass ratio~\cite{Cui-Fermi_Bose}.
However, as pointed out in~\cite{Cui-Fermi_Bose}, $a_{Db}$ may receive large in-medium corrections from three-body effects, which must be calibrated to the system under consideration.
We thus consider $a_{Db}/a_{fb}$ as an unknown parameter which must be carefully measured, and present our main results for a range of plausible values.

\begin{figure}[htbp]
  \includegraphics[width=\columnwidth]{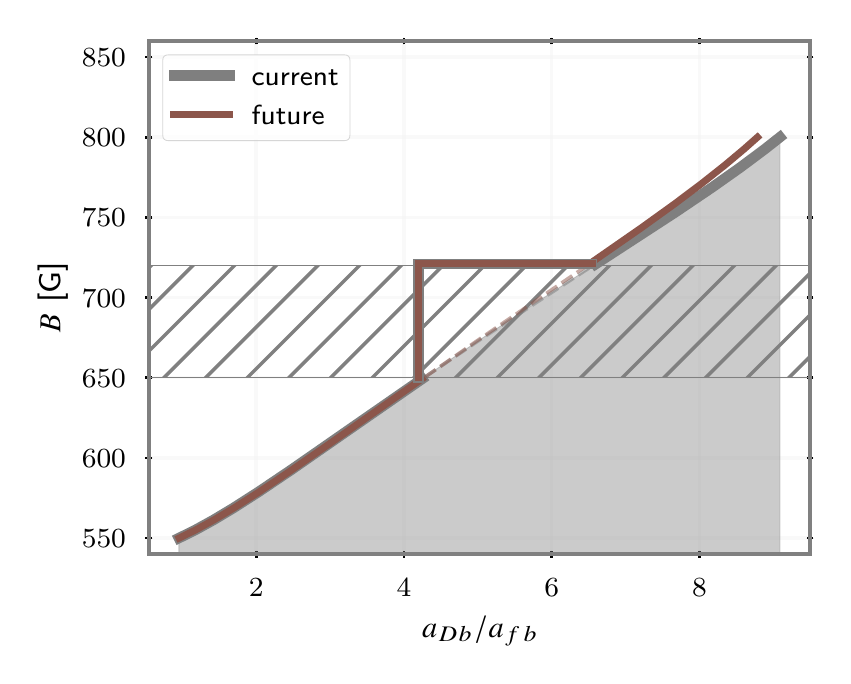}
  \caption{\label{fig:adb}%
    Miscibility condition \cref{eq:miscibility} as a function of $a_{Db}/a_{fb}$~\cite{Green:2019} and magnetic field $B$.
    The shaded region corresponds to immiscible fluids which cannot be used to measure entrainment.
    To maximize the entrainment signal, one should generally choose the smallest magnetic field allowed by the miscibility condition (solid lines).
    Within the hatched region, however, particle losses are particularly high~\cite{Gupta:2012}.
    Thus if $4 \lessapprox a_{Db}/a_{fb} \lessapprox 6.5$, one should keep $B\approx \SI{730}{G}$ to maintain a reasonable lifetime of the system.
    }
\end{figure}

The resulting energy-density functional can be cast in terms of a wavefunction-dependent effective mass matrix for the two-component wavefunction $\vect{\Psi} = (\psi_D, \psi_b)^T$:
\begin{subequations}
  \begin{gather}
    \mathcal{E}[\Psi] = \frac{\hbar^2}{2}\vect{\nabla}\vect{\Psi}^\dagger \cdot\mat{M}^{-1}\cdot \vect{\nabla}\vect{\Psi}
    + \mathcal{E}_h,\\
    \mat{M}^{-1}[\psi_D, \psi_b] = \begin{pmatrix}
      m_D^{-1} - f_{\mathrm{ent}}\frac{m_b}{m_D}n_b
      & f_{\mathrm{ent}}\psi_b^\dagger\psi_D \\
      f_{\mathrm{ent}}\psi_D^\dagger\psi_b & m_b^{-1} - f_{\mathrm{ent}}\frac{m_D}{m_b}n_D
    \end{pmatrix},\label{eq:E}
  \end{gather}
  where
  \begin{gather}
    f_{\mathrm{ent}}(n_D, n_b) = \frac{\rho_{\mathrm{dr}}}{m_Dm_bn_Dn_b}.
  \end{gather}
\end{subequations}
Varying the associated action with respect to $\vect{\Psi}^\dagger$ gives the following coupled non-linear Schr\"odinger equations:
\begin{gather}
  \I \hbar e^{\I\eta}\dot{\vect{\Psi}} = -\frac{\hbar^2}{2} \vect{\nabla} \Big(\mat{M}^{-1}[\psi_{D,b}]\cdot \vect{\nabla} \vect{\Psi}\Big)
  + \mat{V}_{\mathrm{eff}}[\psi_{D,b}]\cdot\vect{\Psi},
  \label{eq:DGPE}
\end{gather}
where we include a complex phase $\eta$ to simulate thermal dissipation (sometimes referred to as the \gls{DGPE} or \gls{PDGPE}) and the effective potential is
\begin{gather*}
  \begin{multlined}
    \vect{V}_{\mathrm{eff}} = \begin{pmatrix}
      \pdiff{\mathcal{E}_{h}}{n_D} & 0\\
      0 & \pdiff{\mathcal{E}_{h}}{n_b}
    \end{pmatrix}
    +
    \alpha
    \frac{\hbar^2}{2}
    \begin{pmatrix}
      \pdiff{f_{\mathrm{ent}}}{n_D} & 0 \\
      0 & \pdiff{f_{\mathrm{ent}}}{n_b}
    \end{pmatrix} +
    \\
    -
    f_{\mathrm{ent}}
    \frac{\hbar^2}{2}
    \begin{pmatrix}
      \frac{m_D}{m_b}\abs{\vect \nabla \psi_b}^2
      & -\vect \nabla \psi_b^\dagger\cdot\vect{\nabla} \psi_D \\
      - \vect \nabla \psi_D^\dagger\cdot\vect{\nabla} \psi_b
      &
      \frac{m_b}{m_D}\abs{\vect \nabla \psi_D}^2
    \end{pmatrix},
  \end{multlined}\\
  \begin{multlined}
    \alpha =
    - \frac{m_b}{m_D}n_b\abs{\vect{\nabla} \psi_D}^2
    - \frac{m_D}{m_b}n_D\abs{\vect{\nabla}\psi_b}^2 + \\
    + \psi_D^\dagger\psi_b\vect{\nabla} \psi_b^\dagger\cdot\vect{\nabla}\psi_D
    + \psi_b^\dagger\psi_D\vect{\nabla}\psi_D^\dagger\cdot\vect{\nabla} \psi_b.
  \end{multlined}
\end{gather*}
\exclude{
\begin{align*}
  \vect{V}_{\mathrm{eff}} =& \begin{pmatrix}
    \pdiff{\mathcal{E}_{h}}{n_D} & 0\\
    0 & \pdiff{\mathcal{E}_{h}}{n_b}
  \end{pmatrix}
  +\\
  &\hspace{-2em}+\begin{pmatrix}
    \alpha \pdiff{f_{\mathrm{ent}}}{n_D}
    - \frac{m_D}{m_b}f_{\mathrm{ent}}\vect \nabla \psi_b^\dagger\cdot\vect \nabla \psi_b
    & f_{\mathrm{ent}}\vect \nabla \psi_b^\dagger\cdot\vect{\nabla} \psi_D \\
      f_{\mathrm{ent}}\vect \nabla \psi_D^\dagger\cdot\vect{\nabla} \psi_b
    & \alpha\pdiff{f_{\mathrm{ent}}}{n_b}
    - \frac{m_b}{m_D}f_{\mathrm{ent}}\vect{\nabla} \psi_D^\dagger\cdot\vect \nabla \psi_D
  \end{pmatrix},\\
  \alpha =&
  - \frac{m_b}{m_D}n_b\vect{\nabla} \psi_D^\dagger\cdot\vect{\nabla} \psi_D
  - \frac{m_D}{m_b}n_D\vect{\nabla}\psi_b^\dagger\cdot\vect{\nabla} \psi_b + \\
  &+\psi_D^\dagger\psi_b\vect{\nabla} \psi_b^\dagger\cdot\vect{\nabla}\psi_D
  + \psi_b^\dagger\psi_D\vect{\nabla}\psi_D^\dagger\cdot\vect{\nabla} \psi_b.
\end{align*}
}

\exclude{
\begin{subequations}
  \label{eq:8}
  \begin{gather}
    \mathcal{E}_h(n_D, n_b) =
    g_{DD}\frac{n^2_D}{2} + g_{bb}\frac{n^2_b}{2}+g_{Db}n_Dn_b,\\
    g_{ii} = \frac{4 \pi \hbar^2 a_{ii}}{m_i}, \qquad
    g_{Db} = \frac{2 \pi \hbar^2 a_{Db}(m_D + m_b)}{m_Dxm_b},
  \end{gather}
\end{subequations}
}

For weakly interacting bosons, $\mathcal{E}_f(2n_D) \equiv g_{DD}n_D^2/2$,
Shevchenko et.\@ al~\cite{shevchenko07} have calculated $\rho_{\mathrm{dr}}$ to leading order in the inter-species coupling $g_{Db}$ in terms of the phonon dispersion relationships $E_{\mathrm{ph}}(\hbar k)$.
We apply their results to our Bose-Fermi mixture by using~\eqref{eq:EphFermi} for the dimers instead of the usual bosonic phonon dispersion.
The resulting expressions are rather complicated to display analytically, so we include a simple numerical implementation in the supplement~\cite{gitlab:entrainment, osf:entrainment}.

This approach captures the correct physics in the deep \gls{BEC} limit of the crossover where $E_{\mathrm{ph}}(\hbar k)$ approaches the usual bosonic dispersion for the dimers, and provides a qualitative extrapolation to the unitary regime.
In the unitary regime, the corresponding expression neglects the softening of the phonon dispersion relationship and pair-breaking effects.
We expect this to provide an upper bound on the magnitude of the entrainment since softening the dispersion \emph{increases} the number of virtual excitations that contribute to entrainment.
This modification of the dispersion relation and hence the enhancement can be achieved by using an optical lattice potential also.
The details of this mechanism are discussed in~\cite{Ol-drag-two-hofer}.

Our approach starts to breakdown in the limit of strong interactions as approximating the interactions using $s$-wave scattering lengths only is not appropriate. Consequently beyond mean-field corrections arise, which should be addressed properly in the calculation.

\section{Proposed Experiment}\label{sec 3}

Using this formalism, we model a mixture of a fermionic (\ce{^{6}Li}, \ce{^2S_{1/2}}) superfluid and a bosonic (\ce{^{174}Yb}, \ce{^1S_0}) superfluid.
Using the wide \textit{s}-wave Feshbach resonance centered at \SI{832}{G}, one can tune the fermionic $a_{ff}$ scattering length between the two lowest hyperfine states of \ce{Li}~\cite{schunck05, nascimbene10}.
Since \ce{Yb} is not magnetically susceptible, the $a_{bb}$ scattering length is fixed.

We propose using an optical ring trap to hold both superfluids with the ring oriented horizontally.
Several convenient values for wavelength exist above \SI{700}{nm}, although the polarizibility of the two species will be different (for instance at a wavelength of \SI{780}{nm} the polarizability ratio of \ce{Li} to \ce{Yb} is about \num{5}), the minima of the potentials coincide, trapping the clouds in the same region of space, but with different trapping frequencies.
In the vertical direction, the weaker gravitational force on the \ce{Li} can be countered with a magnetic field gradient to ensure physical overlap of the superfluids~\cite{Hansen:2013}.

A circular flow around the ring will then be generated in the fermionic superfluid using the procedure of optical stirring~\cite{wright13} with a laser beam which acts as a species selective potential for Lithium. A convenient wavelength is \SI{665}{nm} where the polarizability ratio is about 60.
Such techniques can generate circular flow with 4 windings, corresponding to a flow velocity $v/v_c \approx 0.1$~\cite{wright13} where $v_c$ is the local speed of sound in the center of the cloud.

As discussed below, the entrainment signal is maximized at lower magnetic fields (\gls{BEC} limit) where the fermionic superfluid density increases.
This can however cause two undesirable effects: the immiscibility of the fluids, and increased 3-body loss.
To mitigate these issues, we propose generating the flow at $B=\SI{832}{G}$ \gls{UFG} resonance, then slowly reducing $B$ to maximize the signal once the flow is established.
The loss of miscibility at low $B$ has the interesting side-effect of allowing one to estimate the value of $a_{Db}$ by tuning the magnetic field to the limit of miscibility.


\begin{figure}[htbp]
  \includegraphics[width=\columnwidth]{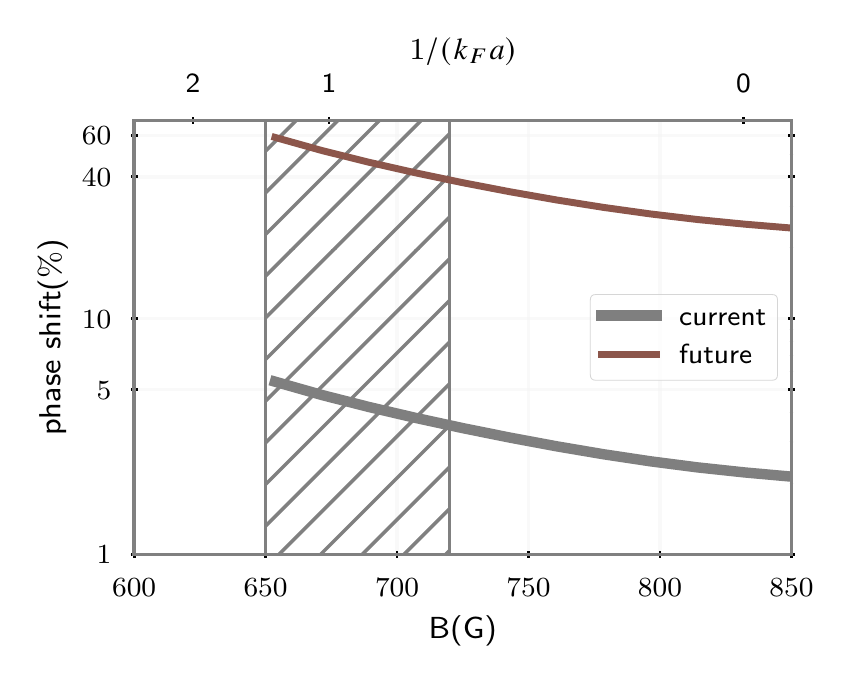}
  \caption{\label{fig:B}%
    Magnetic field dependence of the phase shift for the mean-field value of $a_{Db} = 3.87a_{fb}$.
    The unitary Fermi gas is realized at the resonance $B=\SI{832}{G}$ where $k_Fa_{\ce{Li}} = \infty$.
    Here, we show an example calculation performed for both current and future parameters as listed in \cref{tab:params}.
    The hatched region is the range withing which the particle losses become significantly higher.}
\end{figure}


To prevent flow in the bosonic superfluid, a species-selective repulsive barrier seen only by the bosons can be inserted.
A convenient optical frequency for this potential is about \SI{200}{GHz} blue detuned of the $^{1}S_0$ $\rightarrow$ $^{3}P_1$ transition at \SI{556}{nm} for Yb.
After letting the two superfluids equilibrate, the bosonic superfluid will acquire a phase difference across the barrier due to the entrainment terms.
This can subsequently be measured by observing the interference pattern after expansion~\cite{Andrews:1997}.

This might seem unusual since the barrier arrests any flow in the bosonic superfluid $\vect{v}_b = 0$, so how can a phase shift accumulate?
The resolution is that entrainment modifies the relationship between the phase gradient $\vect{\nabla}{\phi}_{b/D}$ and the group velocity $\vect{v}_{b/D}$:\footnote{In this case, where the momentum operators appear in the functional~\eqref{eq:functional} at only quadratic order, these group velocities can be obtained simply as the derivative of the appropriately defined dispersion: $v_{b,D} = \partial E(\hbar k_b, \hbar k_D)/\partial (\hbar k_{b,D})$.
The presence of higher derivatives complicates this, and one must compute the proper momentum currents in terms of the symmetric ordered expansion of these derivatives.}
\begin{subequations}
  \begin{align}
    \vect{j}_b &=
    n_b\hbar\vect{\nabla}\phi_b
    - \rho_{\mathrm{dr}}\left(
      \overbrace{\frac{\hbar\vect{\nabla}\phi_b}{m_b}}^{\vect{v}_b}
      -
      \overbrace{\frac{\hbar\vect{\nabla} \phi_D}{m_D}}^{\vect{v}_D}
    \right),\\
    \vect{j}_D &=
    n_D\hbar \vect{\nabla}\phi_D
    - \rho_{\mathrm{dr}}\left(
      \frac{\hbar\vect{\nabla}\phi_D}{m_D}
      - \frac{\hbar \vect{\nabla} \phi_b}{m_b}
    \right).
  \end{align}
\end{subequations}
Here, we consider \ce{Yb}-wavefunction (\ce{Li}-dimer wavefunction), $\psi_b =\sqrt{n_b} e^{i\phi_b}$ $(\psi_D = \sqrt{n_D}e^{i\phi_D})$, and $n_b$ $(n_D)$ is the density of the bosonic (dimer) homogeneous state.

The barrier prevents any flow of the bosons, $\vect{j}_{b} = 0$, but the entrainment still allows the accumulation of a phase shift in the presence of fermionic flow $\vect{j}_{D} \neq 0$:
\begin{subequations}
  \begin{align}
  \frac{\hbar\vect{\nabla}\phi_b}{m_b} &= - \frac{\hbar\vect{\nabla}\phi_D}{m_D}\frac{\rho_{\mathrm{dr}}}{m_bn_b}\left(1 - \frac{\rho_{\mathrm{dr}}}{m_b n_b}\right)^{-1},\\
  \vect{j}_{D} &= n_D \hbar\vect{\nabla}\phi_D
  \underbrace{
  \left(1 - \frac{\rho_{\mathrm{dr}}}{m_Dn_D}\left(1 - \frac{\rho_{\mathrm{dr}}}{m_bn_b}\right)^{-1}\right)}_{m_D/m^*_{D}}.
  \end{align}
\end{subequations}
Thus, a circulating fermionic superfluid with $\vect{j}_{D} \neq 0$ will induce a phase gradient $\vect{\nabla}\phi_b \approx -\rho_{\mathrm{dr}}\vect{j}_{D}/(\hbar n_b n_D m_D)$ in the bosons, despite the fact that the barrier keeps the bosonic cloud stationary, $\vect{j}_b = 0$:
\begin{gather}
  \hbar n_b\vect{\nabla}\phi_b
  = -\vect{j}_{D}\frac{\rho_{\mathrm{dr}}}{m_D n_D}
  \left(1 - \frac{\rho_{\mathrm{dr}}}{m_Dn_D}
            \left(1 + \frac{m_Dn_D}{m_bn_b}\right)
  \right)^{-1}.
\end{gather}

Once the phase-shift has been imprinted on the bosons, and the fermions have been removed, the bosonic cloud can be released, allowing the opposite sides of the barrier to expand into each other, forming an interference pattern shifted by the relative phase imprint from entrainment.
Crucial to the success of this protocol is minimizing contamination from mean-field effects, which, as we mentioned in the introduction, can be several orders of magnitude larger than the corresponding entrainment effects, precluding observation through other methods such as the dipole frequency shift in a harmonic trap.
To mitigate this, our setup develops the phase shift in a homogeneous background around the ring geometry.
We have verified through our simulations that the back-reaction from the barrier in the bosonic cloud does not induce any mean-field effects, despite the fermionic flow.
It will be also crucial to remove the fermionic cloud with a sudden vertical laser pulse before performing the interferometery.
Repeating the experiment with circulation in the opposite direction can be used to test for and mitigate any asymmetry in the underlying trap geometry.
Numerical values for the proposed experiment are provided in section~\ref{sec 5}.

\begin{figure}[htbp]
  \includegraphics[width=\columnwidth]{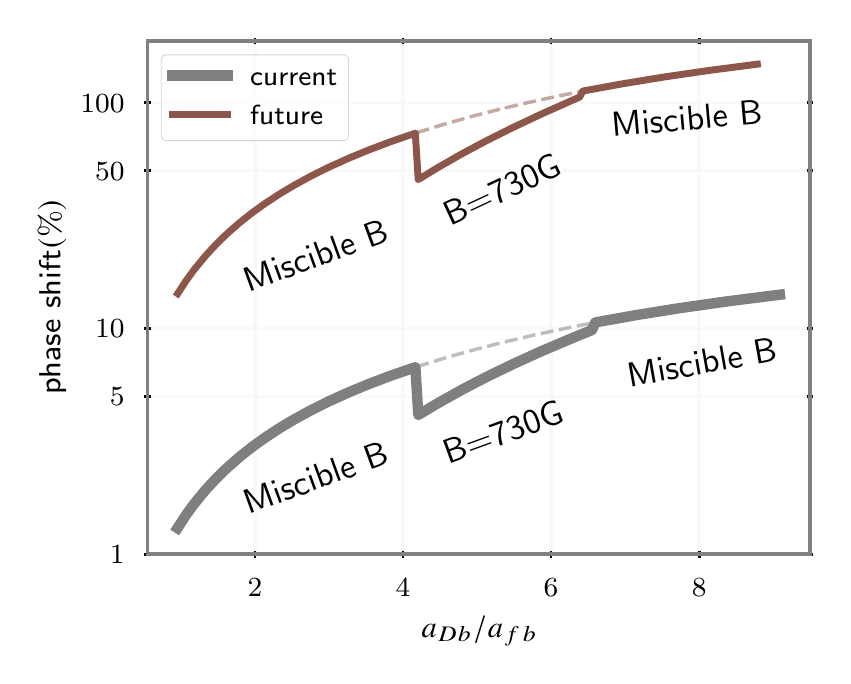}
  \caption{\label{fig:adb_future}%
    Entrainment phase shift ($\Delta \theta/2\pi$) calculated with magnetic field ($B$) values from the miscibility criteria (equation~\ref{eq:miscibility}) for a range of values for $a_{Db}/a_{fb}$.
    The top (bottom) set of curves is for the ``future'' (``current'') values of parameters in table~\ref{tab:params}.
    The dashed region is the high loss region.
    For those $a_{Db}$ values, we recommend $B$ = \SI{730}{G}, for which the mixture remain miscible.
    The phase shift data is plotted in a log scale.
    The points represent the phase shift values calculated using equation~\eqref{eq:scaling} at the fiducial points in table~\ref{tab:params}.}
\end{figure}


\section{Estimated Phase Shift}\label{sec 4}
To estimate the phase shift, we consider a homogeneous gas with wavefunctions:
\begin{align}
  \psi_D(\vect{x}) &= \sqrt{n_D}e^{\I \vect{k}_D\cdot \vect{x}}, &
  \psi_b(\vect{x}) &= \sqrt{n_b}e^{\I \vect{k}_b \cdot \vect{x}}.
\end{align}
with $k_{D,b} = \nabla \phi_{D, b}$.

In our ring geometry, $k_{D} = N_w/R$ where $N_w$ is the number of phase windings, and $R$ is the radius of the ring.
Physically, the magnitude of the winding will be limited by the speed of sound in the superfluid, which defines the local Landau critical velocity $v_c$ which follows from the linearized form of \eqref{eq:EphFermi}: $v_c \approx E_{\mathrm{ph}}(\hbar k)/\hbar k \approx E_{\mathrm{ph}}'(\hbar k)$:
\begin{gather}
  v_D = \frac{\hbar k_D}{m_f} < v_c
  \approx \sqrt{\frac{n_D}{m_D}4\mathcal{E}_f''(n_f)}.
\end{gather}
Minimizing~\eqref{eq:E} we obtain the induced phase gradient in the bosonic component and the corresponding fractional phase shift
\begin{gather}
  \left\lvert\frac{\Delta\theta}{2\pi}\right\rvert = \frac{2\pi R \vect{\nabla}\phi_b}{2\pi}
  =
 \vect{\nabla}\phi_D\frac{R\rho_{\mathrm{dr}}}{m_Dn_b}\left(1 - \frac{\rho_{\mathrm{dr}}}{m_b n_b}\right)^{-1}.
\end{gather}
can be as large as $67\%$, as we show in the next section.

\section{Simulations}\label{sec 5}
We first verify our model, by calculating the dipole oscillation frequency of the bosonic cloud in a \ce{Yb}-\ce{Li} superfluid mixture using the experimental parameters of~\cite{Deep-two-element}.
In this experiment, a small \ce{Yb} \gls{BEC} oscillates in a large \ce{^{6}Li} cloud at $B=\SI{780}{G}$ in a stable mixture of two-superfluids featuring large mass mismatch and distinct electronic properties.
In principle, the entrainment terms should modify the oscillation frequency from that of the background harmonic trap, but as discussed in the introduction, this shift is two orders of magnitude smaller than the shift due to the mean-field interaction.

As validation of our mean-field implementation, we reproduce the observations in~\cite{Deep-two-element}, obtaining a frequency shift of dipole oscillation consistent with the measured values. We include the trap-offset in the direction of the gravity and use $a_{Db} = 2 a_{fb}$.

In the presence of the \ce{Li} cloud, the dipole oscillation frequency ($\omega_{d}$) of the \ce{Yb}-\gls{BEC} is extracted to be $2\pi\,\times\,\SI{381.3(4)}{Hz}$.
From our calculation, the frequency is $2\pi\,\times\,\SI{381.4(9)}{Hz}$ -- well within the experimental errors.
In the experiment, along with the \gls{CM} mode, scissor modes get excited.
The absence of the growth of the scissor mode shows very small energy transfer from the \gls{CM} mode.
Our simulations capture the same qualitative effect.

A subtle point in the experiment was the observation of a decay in the amplitude of the dipole mode.
They conjectured that this may be due to the excitation of quadrupole modes to which the dipole mode is coupled by anharmonicity in the trapping potential.
Within our model, we find that the frequencies of the two modes are sufficiently distinct that energy cannot be efficiently transferred, and suggest instead that this dissipation is due to thermal effects.
We can reproduce the measured decay constants ($\omega_{d}\tau$) within our \gls{DGPE} model \eqref{eq:DGPE} with a phase of $\eta \approx 0.0015$. The measured and calculated values are $250$ and $210$ respectively. They have a reasonable agreement.

Aside from this effect, we reproduce the results with our self-consistent model, verifying the excitation and behavior of scissor modes from misalignment in the trap, and the mean-field frequency shifts.

This validates that our model properly captures the mean-field effects which can potentially obscure entrainment signals.
We have used this model to simulate the proposed experimental procedure to detect entrainment as described in section~\ref{sec 3}, including all mean-field effects and density inhomogeneities due to the trapping potentials.
An important part of this validation is that the induced flow in the fermionic superfluid coupled with the bosonic density perturbation from the barrier \emph{does not} produce an asymmetry in density across the barrier.
This lack of density assymetry differentiates, for example, superfluid entrainment from mutual friction~\cite{Iordansky:1964}: the latter would drag and change the mean-field densities. The main difference is, entrainment interaction couples the gradients of the phases of the two wavefunctions. Therefore, once the transient fluctuations die off, we will have a persistent flow in the fermionic component, which will induce a phase gradient on the bosons leading to a velocity difference across the barrier. Once we remove the barrier in the bosonic component and let the two parts of the cloud expand into each other, the velocity difference will result in a shift in the interference pattern. We may detect this shift by comparing against the interference fringes produced in a system without the induced phase gradient.

\subsection{Experimental Parameters}\label{sec 5a}

We present our results in terms of two sets of parameters.
A \textbf{current} set of parameter values that have been demonstrated through various existing experiments, and a \textbf{future} set of realistic parameter values optimized to detect the entrainment effect.
We find that with current parameters, a modest phase shift occurs which lies at the bounds of current detectability, however, with future improvements, a significant phase shift will be induced that should enable the first direct detection of superfluid entrainment.


\begin{figure*}[htbp]
  \includegraphics[width=\textwidth]{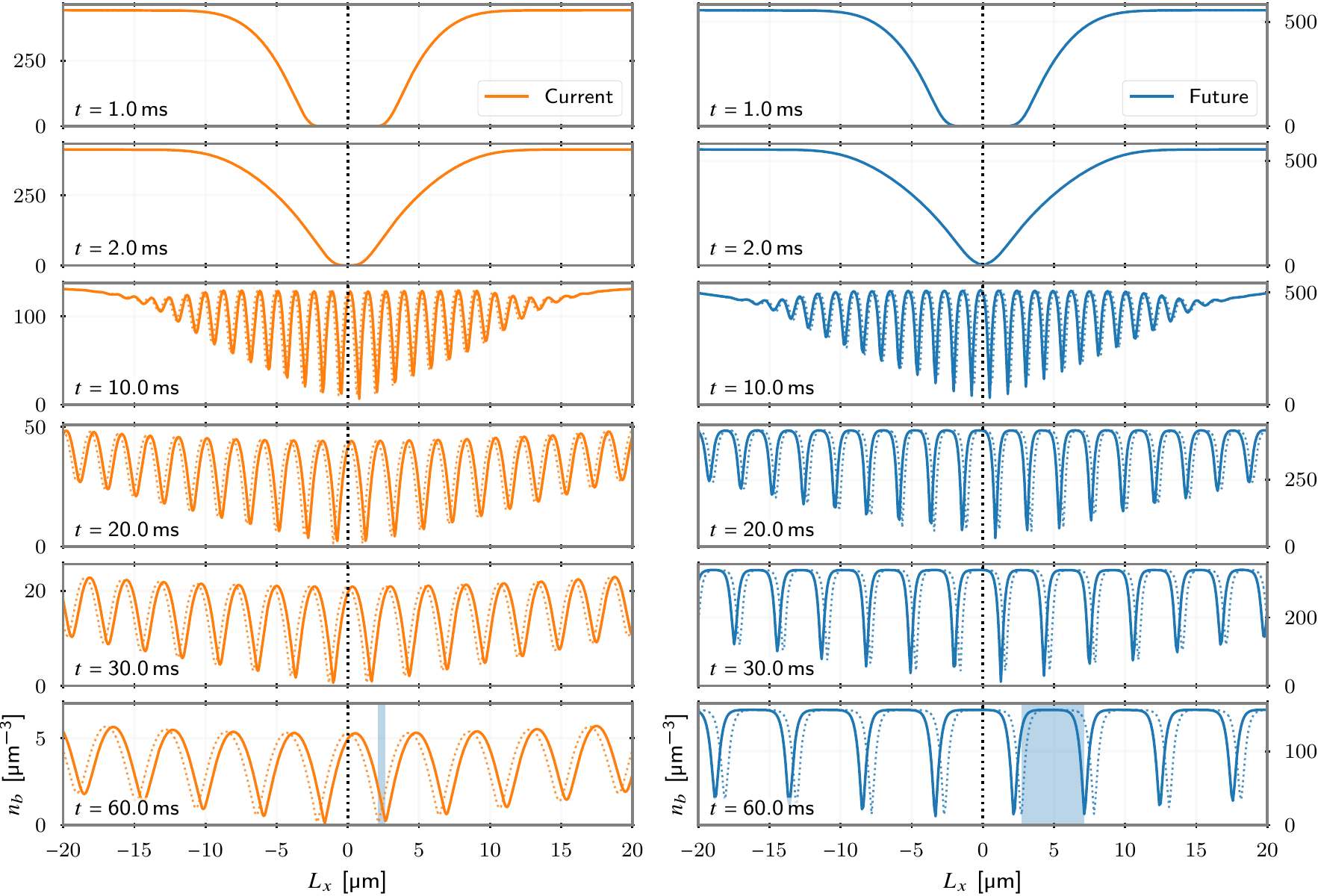}
  \caption{\label{fig:signal}%
    Free expansion of the central density of the bosonic component of the mixture in a periodic box, using the current parameter set (left) and future parameters (right) (see table~\ref{tab:params}).
    Notice the asymmetry around the vertical line at $L_x = 0$: a signature of the measurable entrainment shift.
    These expansion simulations are done with full 3D parameters in a tube geometry which reduces the complexity of the numerical calculations significantly while having the essential aspects of the full 3D calculation.
    Consequently they do not capture the full radial expansion of the ring, but mimic the density drop.
    Back to back measurements will also make the signal more comparable to the expansion of a state without asymmetry.
    In the calculations here, we have used $a_{Db} = 3.87a_{fb}$ and chose the two parts of the cloud to be $\approx \SI{6}{\micro m}$ separated.
    The dotted lines represent the expansion without any entrainment effect, to distinguish the amount of the shift.
    The lightly shaded region in the bottom-right panel shows the amount of shift $\approx \SI{5}{\micro m}$.}
\end{figure*}


\paragraph*{Current Parameters:}

Current experiment in ring traps have demonstrated the ability to trap $N_b \approx 10^5$ bosonic atoms~\cite{Ramanathan:2011}, and recently, $\approx 10^4$ fermionic atoms~\cite{Cai:2021}.
In harmonic traps $N_b = N_f \approx 10^5$ has been achieved in a Fermi-Bose mixture~\cite{Ferrier-Barbut-Bose-Fermi, Deep-two-element} and $N_f \approx 10^6 - 10^7$ has been reported for single species~\cite{Zwierlein:2006}.
With current technology, we expect experimental improvements to trap higher numbers of fermions in the near future.

\begin{table}[htbp]
  \begin{tabular}{r l@{\hskip 5em} r l}
    \toprule
    \multicolumn{2}{l}{Adjustable Parameters} &
    \multicolumn{2}{l}{Fixed Parameters:}\\
    \midrule
    $N_{f}$: & \numrange{e5}{e6}           & $m_{f}=$ & \SI{6}{\atomicmassunit}\\
    $N_{b}$: & \numrange{e5}{e6}           & $m_{b}=$ & \SI{174}{\atomicmassunit}\\
    $B$:     & Miscibility limited         & $\omega_{Yb}=$ & $2\pi\,\times\,$\SI{400}{Hz}\\
    $R$: & \SIrange{30}{200}{\micro m}     & $a_{bb}=$ & \SI{5.5}{nm}\\
         &                                 & $a_{fb}=$ & \SI{1.59}{nm}\\
         &                                 & $\alpha=$ & \num{6.2}\\
    $v/v_c$: & \numrange{0.15}{0.2}        & $\xi=$ & \num{0.3705}\\
    \midrule
    $\Delta\theta/2\pi$: & \SIrange{3}{33}{\percent}\rlap{\qquad ($+$ in \cref{fig:adb_future})} &
    $a_{Db}=$ & $2a_{fb}, \SI{578}{G} $\\
    $\Delta\theta/2\pi$: & \SIrange{6}{67}{\percent}\rlap{\qquad ($\times$ in \cref{fig:adb_future})} &
    $a_{Db}=$ & $3.87a_{fb}, \SI{638}{G}$\\
    \bottomrule
  \end{tabular}
  \caption{\label{tab:params}
    Parameters ranging from current to future values and the corresponding phase-shifts.
  }
\end{table}

The main limit on the maximum density is particle loss due to 3-body processes which scale as $n^3$.
We can mitigate this to some extent by adjusting the fermion density with the magnetic field, and the boson density with the trap frequencies.
For our analysis, we fix the trap frequency for the bosons at $2\pi\,\times\,$\SI{400}{Hz}.  This gives a lifetime for the \ce{Yb}-\gls{BEC} of $\approx$ \SI{2}{s}~\cite{Takasu:2003} with a central density $\approx \SI{340}{\micro m}^{-3} \approx \SI{3.4e14}{cm}^{-3}$ and chemical potential $\mu_b \approx \SI{65}{nK} (\mu_b/\hbar\omega_{Yb}\approx 3)$ in a \SI{30}{\micro m} trap.
Excessively tight traps can heat the system, reducing the expected lifetime.
We propose using the same laser to trap both chemical species, which limits the relative polarizability ($\alpha$) between the two species.
With $\alpha = 6.2$, at $\SI{638}{G}$, the trap frequency for fermions becomes $\approx 2\pi\,\times\,\SI{5}{kHz}$, setting the lifetime of \ce{Li}-dimers at $\approx$\SI{107}{ms} with chemical potential $\mu_D \approx \SI{1.43}{\micro K}$ ($\mu_D/\hbar\omega_{Li}\approx 5)$ and central density $\approx\SI{100}{\micro m}^{-3} \approx \SI{1e14}{cm}^{-3}$.
Here, the \ce{Yb}-\ce{Li} scattering length remains roughly constant.\footnote{Narrow Feshbach resonances ($<\SI{1}{mG}$) have recently been observed in Yb-Li mixtures~\cite{Green:2019feshbach}. These do not affect our discussion.}
These densities are dilute in the sense that mean-field analyses are valid.
Standard values for relevant scattering lengths in the system are: $a_{bb}$, the boson-boson scattering length, \SI{5.5}{nm}~\cite{Kitagawa-Yb} and $a_{fb}$, fermion-boson scattering length, \SI{1.59}{nm}~\cite{Green:2019}


\begin{figure*}[htbp]
\begin{minipage}[b]{\textwidth}
 \includegraphics[width=\textwidth]{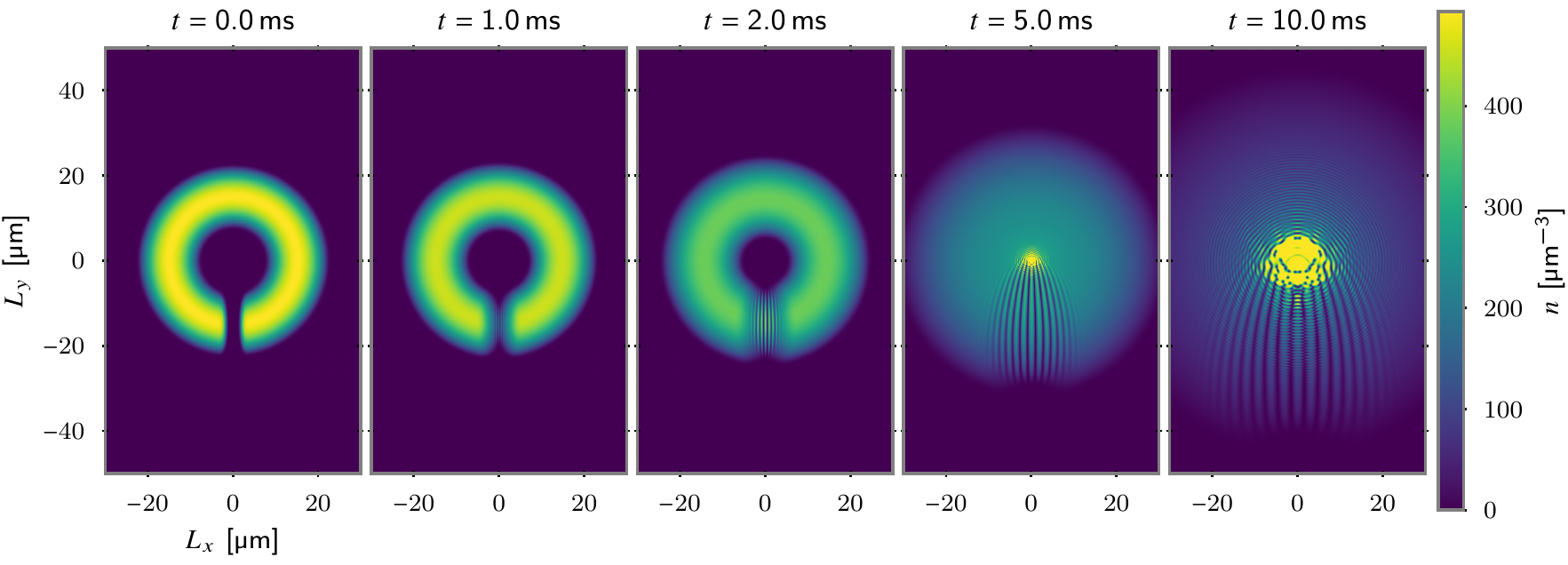}
 \caption{\label{fig:radial}%
   Schematic radial expansion of a ring with an induced phase gradient in the bosonic component.
   As time progresses (left to right), the density decreases because of the outward radial expansion and the fringes become wider.
  (This is schematic in the sense that it is in an effective 2D geometry with translation invarience out of the page, and the visual effect is enhanced by increasing the density.)}
\end{minipage}
\end{figure*}


In table~\ref{tab:params} we compare two sets of adjustable parameters. In the table, $v/v_c$ corresponds to the flow velocity as a fraction of Landau critical velocity in \ce{Li}, $\xi$ is the Bertsch parameter and rest of the parameters have their usual meaning. The left-hand side of the first column shows typical values realized in current experiments. The right-hand side shows how these could be adjusted to maximize the entrainment signal with values that should be attainable with reasonable experimental advances.

We have calculated the percentage phase shift from the unitarity up to the \gls{BEC} side of the phase diagram for the Fermi-Bose mixture by numerically integrating the equation (28) of ~\cite{fil05} at T = 0 with modified equation of state and phonon dispersion relationship for the fermionic dimers. We have also determined the parameter dependencies of the percentage phase shift from scaling analysis, and it has the following form:
%
\begin{multline}\label{eq:scaling}
\frac{\Delta \theta}{2\pi} \approx
\Bigl(\SI{6.2}{\percent}\Bigr)
\Big(\frac{v}{0.15v_c}\Big)
\Big(\frac{a_{Db}}{3.87a_{fb}}\Big)^{\num{2.02(1)}}
\Big(\frac{B}{\SI{638}{G}}\Big)^{\num{-3.4(3)}}\\%
\Bigl(\frac{\alpha}{6.2}\Bigr)^{\num{0.58(1)}}
\Big(\frac{R}{\SI{30}{\micro m}}\Big)^{\num{0.46(1)}}
\Big(\frac{N_f}{10^5}\Big)^{\num{0.58(1)}}
\Big(\frac{N_b}{10^5}\Big)^{\num{-0.033(1)}}\\%
\Big(\frac{m_b m_f}{m_{Yb}m_{Li}}\Big)^{\num{0.54(1)}}
\Big(\frac{m_b/m_f}{m_{Yb}/m_{Li}}\Big)^{\num{0.02(1)}}.
\end{multline}
%
The exponents in the equation \eqref{eq:scaling} demonstrate the approximate dependence of the phase shift on various parameters.
The errors in the exponents demonstrate how the exponents vary due to the full functional dependence provided in the accompanying code~\cite{gitlab:entrainment, osf:entrainment}.
accounts for the changes in values for the currently achievable parameters (left column in \cref{tab:params}) from the possible experimental parameters implementable in future (right column in \cref{tab:params}).
See~\ref{tab:errors} for a complete tabulation of these exponents and their errors.

\paragraph*{Future Parameters:}

From \cref{eq:scaling} we see that the most significant signal enhancements come from increasing the dimer-boson scattering length $a_{Db}$ and decreasing the magnetic field $B$ towards the \gls{BEC} limit.
Miscibility limits the enhancement of these two effects, requiring one to adjusting the magnetic field $B$ according to \cref{eq:miscibility}.
To maximize the entrainment signal, one should use the lowest magnetic field allowed by miscibility (see \cref{fig:adb}), but particle loss rates are particularly high in the range \SIrange{650}{720}{G}~\cite{Gupta:2012}.
At each magnetic field, the miscibility condition gives us values of $a_{Db}$ for which the mixture remains miscible. If the unknown $a_{Db}$ value lies within the miscible $a_{Db}$ range of \SIrange{650}{720}{G}, then one should keep $B \approx \SI{730}{G}$ to minimize particle loss.

Other parameters that have positive effect on the enhancement are the radius of the ring ($R$), and the induced flow velocity ($v/v_c$).
In general, by increasing each of them we can increase the signal strength, but each of them is limited by potential experimental challenges.
Perfect larger ring traps are harder to obtain as they may not remain flat at the center along the azimuthal axis.
This will lead to production of higher amount of mean-field density fluctuation which
may not completely die out within the lifetime of the metastable superfluid state.
For a single bosonic species $R \approx \SI{250}{\micro m}$~\cite{Sherlock:2011} has been achieved.
This order of magnitude may be achievable for a superfluid mixture.
In our experimental protocol, creating a substantial flow in the fermionic component increases the signal strength by a considerable amount, but is strictly limited by the number of windings we can induce set by Landau criterion.
There is also the possibility to make an improvement by increasing the induced velocity to as close as possible to the critical velocity.
Typically, only $\approx 10\%$ of the critical velocity is accessed~\cite{Ryu:2007}.
A key part of our protocol is to used the same ring trap for both species which will ensure they are trapped in the same physical space.
The centrifugal force of the rotating fluid displaces the fermionic cloud slightly, but with these parameters the displacement is insignificant $\lesssim \SI{0.014}{\micro m}$.

With the current parameter set, we may produce a phase shift up to $\approx 6\%$ (considering the mean-field enhancement of $a_{Db}$) of observable phase shift. The suggested improvements (future parameters) may increase the signal to as high as $\approx 67\%$.

\subsection{Entrainment Signal}\label{sec 5b}

We present snapshots from the 1D time evolution of the bosonic component in a tube with the current and future parameters in figure~\ref{fig:signal}.
These are time evolution images, which demonstrate the asymmetry in the interference fringes generated from the entrainment interaction.
In an actual experimental procedure, the interfering clouds will be imaged after a time of flight expansion after turning off the confining potential. Where the density will drop and the mean-field interaction between the bosons will reduce.
In that scenario, the fringe spacing after a long expansion time $t$ is:
\begin{equation}
\lambda_s \approx \frac{2\pi \hbar t}{m d}
\end{equation}
for initial separation $d$, as seen in the earlier experiments~\cite{Andrews:1997}.
Essentially, with long time-of-flight expansion, the fringe spacing increases linearly with time and $\SI{30}{\milli s}$ is typical for earth bound laboratory experiments. This can be increased by a factor of $3$ to $10$ in microgravity setting~\cite{van-Zoest:2010} as the reduced gravity will substantially increase the time-of-flight.
By employing a linear optical potential (with a laser far detuned from the Yb-resonances), one may also reduce the effect of gravity and increase the time of flight~\cite{Shibata:2020}.
Competing factors here are the initial separation and the mean-field interaction. The clouds have to be initially separated enough that, they can grow large in size before the interference.
Therefore, a point of suitable compromise has to be reached to maximize the signal in a typical laboratory experiment.
The reduction in density during the time-of-flight has another positive effect on the experimental outcome. This widens the fringes, which will make them easier to resolve.
To demonstrate this effect we perform simulations in a tube geometry, which mimics the density reduction in a radially expanding cloud. Similar effect is also visibly present in the 2D expansion of the ring in figure~\ref{fig:radial}.

\section{Conclusion}\label{sec 6}
We propose an experimental protocol to directly measure for the bulk 3D entrainment -- the Andreev-Bashkin effect -- in a fermionic and bosonic superfluid mixture.
By using a ring geometry with a common trap for both species, we demonstrate that atom interferometry techniques can produce measurable phase shifts with reasonable experimental parameters, and we characterize their dependence.
An important feature of our protocol is the elimination of mean-field effects which can thwart previous attempts to measure entrainment based on shifts in the dipole oscillation frequency in a harmonic trap.
The broad Feshbach resonance allows us to adjust the interaction strength of the \ce{^{6}Li} superfluid to ensure miscibility, and the large mass of the bosonic \ce{^{174}Yb} superfluid enhances the entrainment effect.
One uncertainty is the dimer-boson scattering length $a_{Db}$, which may have significant corrections from three-body physics~\cite{Cui-Fermi_Bose}.
The value of this parameter still needs careful measurement, but a significant enhancement to the entrainment signal is possible if this is large, raising an intriguing potential for developing a significant entrainment effect by tuning three-body interactions.
With current technologies and assuming mean-field values for $a_{Db}$ a \SI{6}{\percent} phase shift should be reliable.
With reasonable improvements in experimental techniques, this may be as larger as \SI{67}{\percent}.

\section{Appendix}

\subsection{Radial Expansion}
Here we demonstrate snapshots from a schematic radial expansion of a ring in 2D in figure~\ref{fig:radial}. This simulation is done with the current parameters in table~\ref{tab:params}, except we modified the density, and the width of the ring to enhance the interference effect allowing more visibility for the demonstration purpose. This demonstration reiterates the importance of time-of-flight expansion in the detection procedure. Longer time-of-flight expansion make the fringes easier to resolve by increasing the fringe width.

\subsection{Deviations in the exponents}
Table~\ref{tab:errors} shows the deviation in the exponents that appear in~\cref{eq:scaling}.

\begin{table}[htbp]
  \begin{tabular}{lll}
    \toprule
    Parameter & Current exponent & Future exponent\\
    \midrule
    $\alpha/\num{6.2}$ & \num{0.59203+-0.00002} & \num{0.56486+-0.00003}\\
    $N_b/10^5$ & $\num{-0.031+-0.003}$ & $\num{-0.034+-0.003}$\\
    $N_f/10^5$ & $\num{0.59202+-0.00002}$ & $\num{0.56486+-0.00003}$\\
    $(m_b m_f)/(m_{Yb}m_{Li})$ & $\num{0.5609+-0.0001}$ & $\num{0.5308+-0.0001}$\\
    $R/\SI{30}{\micro m}$ & $\num{0.4391+-0.0001}$ &$\num{0.4691+-0.0001}$\\
    $B/\SI{638}{G}$ & $\num{-3.18+-0.05}$ & $\num{-3.67+-0.02}$\\
    $a_{Db}/3.87a_{fb}$ & $\num{2.0152+-0.0002}$ & $\num{2.0279+-0.0003}$\\
    $(m_b/m_f)/(m_{Yb}/m_{Li})$ & $\num{0.02+-0.01}$ & $\num{0.03+-0.01}$\\
    \bottomrule
  \end{tabular}
  \caption{\label{tab:errors}%
    Extracted parameter dependence for current and future parameter values.
    These are the powers $p$ appearing in corresponding terms of \cref{eq:scaling}.
    The errors express the typical range over which the exponents vary when the parameters are changed by $\pm \SI{10}{\percent}$ about the current or future values listed in \cref{tab:params}.
  }
\end{table}

\subsection{Equation of state: \textbf{\ce{^{6}Li}}}

To perform our calculations, we parametrize the equation of state of $^{6}$\ce{Li} Fermi gas.
Once we fix the scattering length ($a_{\ce{Li}}$), in terms of the external magnetic field ($B$), we use the following model:
\begin{subequations}
\begin{gather}
\mathcal{E}(n, x) = f(x)\mathcal{E}_{FG}(n)\\
f(x) = \frac{bx + \xi}{\frac{x}{\xi}(b+\zeta) + 1 + \frac{18\pi b x^2}{5 a_{DD}/a_{\ce{Li}}}}
\end{gather}
\end{subequations}
with $x = 1/(k_F a)$, $\mathcal{E}_{FG}(n) = \frac{3}{5}n \hbar^2 k_F^2/2m_F$.
This Padé expansion has two following limits:
\begin{gather*}
  \frac{\mathcal{E}(n)}{\mathcal{E}_{FG}(n)} = \begin{cases}
    \xi - \zeta x + \order(x^2) & x \approx 0 \; \text{(Unitarity)}.\\
   -\frac{5}{3}x^2 + \frac{a_{DD}}{a_{\ce{Li}}}\frac{5}{18\pi x} + \order(\frac{1}{x^{2}}), & x \rightarrow \infty \; \text{(BEC)}.
  \end{cases}
\end{gather*}
Here, $a_{DD}$ is the dimer-dimer scattering length, $\zeta$ is the contact, and $\xi$ is the Bertsch parameter.
\begin{align*}
  \xi &= \num{0.3742(5)}, & \zeta &\approx \num{0.901}, & a_{DD} &= 0.6 a_{\ce{Li}}.
\end{align*}
The more relevant parameter in this case, the dimer-dimer scattering length, we calculate it using $a_{DD} = 0.6a_{Li}$~\cite{Petrov-Fermion-Dimer}.

In the \gls{BEC} limit, this reproduces the expected equation of state for a BEC of dimers with density $n_D = n/2$, mass $m_D=2m_F$, and scattering length $a_{DD}$.
The value of $b = 0.25$ is chosen to match the \gls{QMC} data on the \gls{BEC} side of the phase diagram.
\begin{align*}
  \mathcal{E}_{BEC}(n) &= -\frac{\hbar^2}{2m_F a_{\ce{Li}}^2}n + \frac{g_{DD} n_D^2}{2} + \cdots, &
  g_{DD} &= \frac{4\pi \hbar^2 a_{DD}}{m_D}
\end{align*}
To relate the magnetic field strength $B$ with the scattering length, we fit the data from~\cite{Zurn2013}, using the model from~\cite{Bartenstein:2005}:
\begin{gather}
  a_{\ce{Li}}^{-1} = a_{bg}^{-1}\frac{B-B_0}{(B - B_0 + \Delta) (1 + \alpha(B-B_0))}\\
  B_0 = \SI{832.178498}{G}, \quad
  \Delta = \SI{293.396620}{G},\nonumber \\
  a_{bg} = \num{-1415.05108}~a_B, \quad
  \alpha = \num{0.000406405370}, \nonumber \\
  a_B=\SI{0.0529177}{nm}.\nonumber
\end{gather}
\subsection{Entrainment Coefficient}

We have used equation (28) of ~\cite{fil05} at T = 0 as $\rho_{\mathrm{dr}}$ in our calculation. The analytic form is the following:
\begin{equation}
\rho_{\mathrm{dr}} = \int \frac{dk}{2\pi^2} k^2 \gamma_{Db}^2 \sqrt{m_Dm_b}\frac{n_D n_b \left(\epsilon_{D}\epsilon_{b}\right)^{3/2}}{\Omega_{D}\Omega_{b}\left(\Omega_{D}+\Omega_{b}\right)^{3}}.
\end{equation}
with
\begin{gather*}
\epsilon_{i} = \frac{\hbar^2 k^2}{2m_i}, \qquad E_i = \left(\epsilon_i(\epsilon_i + 2 g_i n_i)\right), \\
\gamma_{Db} = \frac{2\pi\hbar^2a_{Db}(m_D+m_b)}{m_Dm_b}, \qquad g_{bb} = \frac{4\pi\hbar^2a_{bb}}{m_b}, \\
g_{DD} = \frac{\partial^2\mathcal{E}_{f}(2n_{D})}{\partial n_D^2}.
\end{gather*}
and
\begin{equation*}
\Omega_i = \left(\frac{E_b^2 + E_D^2}{2} \pm \sqrt{\frac{(E_b^2 - E_D^2)^2}{4} + 4 \gamma_{Db}^2 n_D n_b \epsilon_{D}\epsilon_{b}}\right)^{1/2}.
\end{equation*}
We have numerically verified that, in the regions of experimental interest,
\begin{equation}
\rho_{\mathrm{dr}} \approx \int \frac{dk}{2\pi^2} k^2 \gamma_{Db}^2 \sqrt{m_Dm_b}\frac{n_D n_b}{\Omega_D\Omega_b}.
\end{equation}
follows similar scaling relationships as $\rho_{\mathrm{dr}}$, and can be used to intuitively understand the scaling relationship shown in equation~\eqref{eq:scaling}.
The essence is that, in the integral, the complicated cubic dependencies essentially drop out and we can use the simplified relationship for qualitative discussions and rough estimates of the scaling exponents.
For quantitative results, we numerically integrate $\rho_{\mathrm{dr}}$ in the thermodynamic limit for homogeneous matter with all dependencies.
To estimate the finite-size effects from the confinement of the ring, we computed the integrals as momentum sums in a finite periodic box, and find only a small correction of \SIrange{1.86}{2.5}{\%} for future and current parameters respectively.
For completeness, we write the explicit parameter dependencies below:
\begin{align*}
  &\gamma_{Db}(a_{Db}, m_D, m_b),
  & &n_D(N_f, m_f, R, \omega_f, \xi, B) \\
  &n_b(N_b, m_b, \omega_b, a_{bb}, R),
  & &\omega_f(\alpha, \omega_b).
\end{align*}
\vspace{0.1em}
\section*{Acknowledgements}
We acknowledge the support of National Science Foundation (NSF) under grant number 1707691 and 1306647. K.H. is grateful to Jethin Pulikkottil Jacob for many insightful discussions.






\providecommand{\selectlanguage}[1]{}
\renewcommand{\selectlanguage}[1]{}

\bibliography{master,local}

\end{document}